\documentclass[twocolumn,aps,showpacs,prl,floatfix]{revtex4}
\usepackage{epsfig,amsmath,amssymb}

\def\be{\begin{equation}} 
\def\ee{\end{equation}} 
\def\bee{\begin{eqnarray}} 
\def\eee{\end{eqnarray}}

\begin{document}

\title{Memory-effect in glasses at low temperatures} 
\author{Peter Nalbach}
\affiliation{Department of Physics, Stanford University, Stanford, CA
94305, USA} 
\date{\today}

\begin{abstract}
The dielectric constant of amorphous solids 
at low temperatures is governed by the dynamics of tunneling
systems, small 
groups of atoms which tunnel between quasi equivalent potential
minima. Recent experiments \cite{Danna02} showed that at temperatures below
$20$~mK various glasses exhibit memory for a previously applied electric bias
field. A first sweep of an electric bias field may prepare
resonant pairs of tunneling systems, which are temporarily formed during the
sweep, in metastable states. In subsequent sweeps the same resonant pairs thus
significantly contribute to the dielectric constant, leading to a higher
dielectric constant. We investigate the dynamics of resonant pairs during a
bias field sweep yielding a qualitative explanation of the memory effect.
\end{abstract}



\date{\today}
\maketitle

Glasses at low temperatures exhibit properties which differ considerably from
those of their crystalline counterparts \cite{ZellerPohl71}.
They are thought to be caused by tunneling
systems (TSs) stemming from the tunneling motion of small atomic entities in
double-well potentials. The energy splitting $E$ of these TSs is given by
$E^2=\Delta^2+\Delta_0^2$, where $\Delta$ is the asymmetry energy due
to the difference in the depth of the two wells and $\Delta_0$ is the
tunneling splitting originating from the overlap of the wave
functions. The randomness of the glassy structure results in broad
distributions for these two parameters.
The phenomenological tunneling model
successfully describes most of the thermal, elastic and dielectric properties
of glasses at low temperatures \cite{Phillips}.
However, experiments below $100$~mK
have revealed that deviations from the predictions of the tunneling model
exist. Although a comprehensive theory is still missing, it seems that the
interaction between the TSs gives rise to many of these deviations
\cite{Doug}. 

\begin{figure}[th]
\epsfig{file=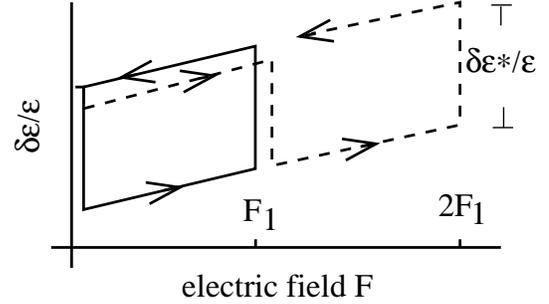,width=7cm}
\caption{\label{fig1}The change of the dielectric constant is plotted over the
  electric bias 
  field. The arrows show the behavior in time. In a first sweep the field is
  swept in a triangle wave between zero and $F_{1}$. The full line
  illustrates the change of the dielectric constant during the first
  sweep. The dashed line 
  represent the behavior during the second sweep where the maximum field is
  twice as large as in the first sweep but the sweep rate is the same.}
\end{figure}

In recent experiments D. Rosenberg et al. investigated the AC dielectric 
constant during the first few sweeps of an electric bias field after the
sample was cooled from about $1$~K to about $10$~mK \cite{Danna02}. 
The electric bias field was swept in
triangular waves for two successive sweeps. The electric bias field amplitude
of the second 
sweep was twice as large as the amplitude $F_1$ of the first but
the sweep rate ${\rm d}F/{\rm d}t=F_{1}/t_{R}$ with the period $2t_{R}$ of the
first triangular wave, was kept constant. 
At the beginning of the first sweep the dielectric constant suddenly decreased
but then increased slowly with increasing electric field. When the field
reached the maximum and started to decrease again the
dielectric constant suddenly increased and then decreased slowly with 
decreasing field to about its original value when the field reached
$F=0$. The solid line in Fig. \ref{fig1} illustrates the dielectric constant
during the bias field sweep. 
The slow increase/decrease of the dielectric constant while
increasing/decreasing the bias field follows the dipole-gap behavior found in
earlier experiments \cite{Doug}. 
However, the sudden changes of the dielectric constant at $F=0$
and at $F=F_1$ are remarkable new features. Typically the relative change of
the dielectric constant of these sudden changes was
$\delta\varepsilon'^{\star}/\varepsilon'\sim 10^5$. It decreased inversely
proportional to the temperature between $T=5$~mK and $T=20$~mK, where the
effect could not be resolved anymore. $\delta\varepsilon'^{\star}/\varepsilon'$
also increased with increasing sweep rate. 
The most remarkable feature yet happened when the second sweep with twice the
amplitude was performed. 
The dielectric constant smoothly 
increased with increasing field until $F=F_1$. 
At this point the dielectric constant suddenly decreased to the value it had
when this field was applied during the first sweep. 
Further increasing the field, the dielectric constant increased
following the course which we would have extrapolated
from the first sweep. The dashed line in Fig. \ref{fig1} illustrates the
dielectric constant during the second sweep. 
Reaching the maximum field $2F_1$ the dielectric constant suddenly
increased and then decreased slowly with decreasing field. 
This behavior suggests that the sample remembers the field strength 
which was previously applied to it. 
Only if the sample was left at zero bias field for several days
or warmed above $1$~K would the memory as described vanish.

An electric field $\vec{F}$ couples to the dipole moment $\vec{p}$ of a TS,
thus, changing the asymmetry energy $\Delta(\vec{F})=\Delta+\vec{p}\vec{F}$. 
Since the distribution of asymmetries is flat over an energy range exceeding
typical values of $\vec{p}\vec{F}$ we expect in thermal equilibrium no changes
of the dielectric constant with applied field.
However, we expect the changing field to drive TSs out of equilibrium.
The TSs, which dominantly contribute to the dielectric constant, have
relaxation times $\tau$ typically shorter than 
$\tau=1$~s at the temperatures of interest~\cite{Relaxrate}. 
Accordingly, isolated TSs driven out of equilibrium cannot be responsible for
the observed long time effects. We propose that weak interactions between
TSs cause the memory effect. 

\begin{figure}[th]
  \vspace*{5mm}\epsfig{file=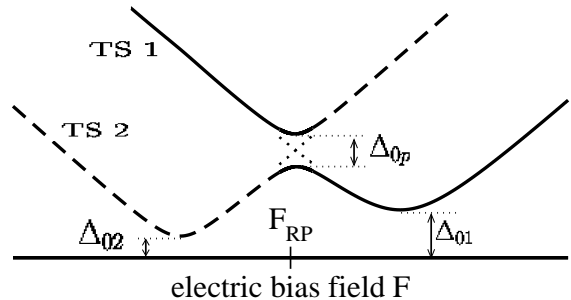,width=7.5cm}
\caption{\label{fig2}The spectrum of two tunneling systems is plotted
  versus bias field. At $F=F_{RP}$ these tunneling systems
  form a resonant pair.}
\end{figure}

Electrically allowed transitions lead to a dielectric response inversely
proportional to the energy splitting and proportional to the occupation
difference between the two states. 
Thus, small energy splittings result in large contributions to the dielectric
constant. On the other hand, small energy splittings lead to almost identical
occupation numbers for the two states so that their response is
suppressed~\footnote{There is an 
  ensemble of TSs for each set of parameters $\Delta_0$ and $\Delta$. By using
  the phrase 'occupation number' for a single TS we refer to the ensemble of
  systems with the same parameters.}. 
However, a field sweep drives the TSs out of equilibrium.  
If two TSs have the same energy splitting $E_1=E_2$ for a specific field
$F_{RP}$ a weak interaction $J_{12}\ll E_1, E_2$ between these two TSs lifts
their degeneracy resulting in a small splitting $\Delta_{0p}\le J_{12}$. 
We illustrate this 'avoided crossing' in Fig. \ref{fig2} where we plot
the excited energy levels of two TSs versus bias field. 
With increasing field the
excitation energy of TS 1 (solid line) decreases until it reaches its minimum
$\Delta_{01}$ and then increases again. Note that the energy splitting
of a TS varies with the electric field since the field shifts its asymmetry. 
TS 2 (dashed line) behaves the same. The levels cross at $F=F_{RP}$ where
these two TSs form a resonant pair (RP). 
The TSs forming the relevant RPs have energy splittings $E\simeq 2k_{\rm B}T$
and in thermal equilibrium the occupation difference of the
split levels is approximately proportional to $\Delta_{0p}/(k_{\rm B}T)$. 
Regarding the small number of RP (about one per $200$ TSs), the excess
dielectric response is negligible. However, we will show later that TSs
with long relaxation times form RPs during the field sweep which are far from 
thermal equilibrium and provide an excess dielectric response of the order of
the observed effect.  

In order to understand how RPs yield memory for the previously applied bias
field consider two TSs forming a RP where TS 1 starts
in the ground state and TS 2 in the excited state. An adiabatic field sweep
from $F<F_{RP}$ to $F>F_{RP}$ results in a flip-flop leaving TS
1 excited and TS 2 in its ground state (compare Fig. \ref{fig2}). 
For an ensemble of such TSs the subensemble of TSs 1 interchanges the 
occupation number of its excited state with that of the ensemble of TSs 2. 
We further assume that the relaxation time of TS 1 is short, so that it
adjusts its occupation number (of its excited state) according to thermal
equilibrium to its momentary energy splitting, and the relaxation time of TS 2
is long, so its occupation number stays unaltered during the experiment. 
The flip-flop exchanges the occupation numbers and TS 2 will store the
equilibrium occupation number to field $F=F_{RP}$.
When the field reaches $F=F_{RP}$ the second time, after it was swept to its
maximum value and then decreased, TS 1 has again an equilibrium occupation
number to the applied field $F=F_{RP}$. Now both TSs are in thermal
equilibrium. 
All further flip-flops store the equilibrium occupation number into TS 2 and
the situation stays unchanged. 
Only during the very first formation of a RP one TS is in thermal
equilibrium. This produces the observed memory effect.

In the following we quantify the above picture and estimate the temperature
and sweep rate dependences.  
We start by calculating the resonant contribution \footnote{At the
  temperatures $T\lesssim 20$~mk 
  relaxational contributions to the dielectric constant are expected to be
  negligible.} to the dielectric response function of a RP
\[ \chi(t-t') \,=\, \langle[P(t),P(t')]\rangle\Theta(t-t') \; . \]
Here $\langle\cdots\rangle$ indicates the thermal and ensemble averages and
$[\cdot ,\cdot ]$ the commutator. $\Theta$ is the Heaviside step function and
$P=p_1\sigma_{z1}+p_2\sigma_{z2}$ the total dipole moment of the two TSs. The
absolute value of $P$ depends on the magnitude and the orientation of
the individual dipoles $p_i$ with respect to the external measuring field. 
The Hamiltonian of a coupled pair of TSs is given by
\[\label{eq1} H \,=\, \frac{\Delta_{01}}{2}\sigma_{x1} \,+
\frac{\Delta_{1}}{2}\sigma_{z1} \, +\frac{\Delta_{02}}{2}\sigma_{x2} \,+
\frac{\Delta_{2}}{2}\sigma_{z2} \,+ J_{12}\sigma_{z1}\sigma_{z2} 
\]
with the Pauli matrices $\sigma_{\alpha j}$ describing the two state variables
of the two TSs.
Considering only lowest order contributions in the interaction we obtain for 
the excess response $\chi_{RP}$ of a RP
\be\label{eq3} \chi_{RP} \,=\,
\left( p_1\frac{\Delta_1}{E_1}+p_2\frac{\Delta_2}{E_2} \right)^2
\left(\frac{\Delta_{0p}}{E_p} \right)^2 \Big( n_+-n_-\Big) 2\sin\Big( E_pt
\Big)  
\ee
with the
occupation differences $n_+=\tanh(E_+/(2k_BT))$ between the ground state and
the second excited state of the RP and $n_-=\tanh(E_-/(2k_BT))$ between the
ground state and the first excited state. The energies 
$E_\pm=(E_1+E_2)/2\pm E_p$ with $E_p=\sqrt{\Delta_{0p}^2+(E_1-E_2)^2/4}$
are the excitation energies of the RP and
$\Delta_{0p}=(\Delta_{01}/E_1)(\Delta_{02}/E_2)J_{12}$ (see Fig.
\ref{fig2}).    
Note that especially asymmetric TSs, $\Delta_i\sim E_i$, forming RPs, have an
excess dielectric response whereas isolated TS only contribute substantially
to the resonant contribution of the dielectric constant when they are
symmetric, $\Delta\ll\Delta_0$. 

At first we estimate the response of RPs in thermal equilibrium. The
standard tunneling model assumes a distribution function
$P(\Delta,\Delta_0)=P_0/\Delta_0$ leading to a flat distribution of energy
splittings. One further assumes a maximum splitting $E_{max}\simeq 1$~K. 
The number $\#n$ of relevant RPs, formed by two thermal TSs,
$E_i\simeq 2k_BT$, where the coupling fullfils $2\Delta_{0p}>|E_i-E_j|$, is
$\#n=(P_02k_BT)\cdot (P_02\Delta_{0p})/(P_0E_{max})$.

A common estimate for the mean coupling between TSs are $\overline{J}\simeq
10~k_B$mK \cite{BurinRelax94}.
Due to the $1/r^3$-dependence of the interaction on the
distance $r$ between the TSs the mean interaction is proportional to the
density of TSs and thus, the mean interaction of thermal TSs is
$\overline{J}(T)\approx 100~k_B\mu$K at $T=10$~mK with $E_{max}=1$~K. 

Assuming further $\Delta_i\sim\Delta_{0i}$ for the TSs forming RPs and $E_p\ll
E_1, E_2$ we roughly estimate the
real part of the dielectric constant (given as the complex Fourier transform
of the response function) for RPs ($\Delta_{0p}\simeq E_p$) at $T=10$~mK to be
\[ \chi'^{(eq)}_{RP} \,\sim\, 4\cdot 10^{-4}\cdot P_0p^2   
\]
with the mean dipole moment $p$. $\chi'^{(eq)}_{RP}$ is negligible compared to
the dielectric response, about $\sim P_0p^2$, of isolated TSs in the tunneling
model. 

During a field sweep most TSs will not be in thermal equilibrium and we have
to investigate the occupation numbers of two TSs which form a
RP at a field $F_{RP}$ during the sweep. For fields $F\not=F_{RP}$ the
coupling between the two TSs is irrelevant. 
Assuming a typical dipole moment of $1$ Debye, typical bias fields
$F_1=133$~kV/m change the asymmetry by an amount corresponding to a
temperature $\Phi/k_B=\vec{p}\vec{F}_{1}/k_B\simeq 30$~mK.
With sweep times $t_{R}=10$~s the time, during which two TSs fullfil the RP
condition $2\Delta_{0p}>|E_i-E_j|$ during the sweep, is 
$33$~ms at $T=10$~mK. This time is short compared to relaxation
times of thermal TSs.
Accordingly we approximate the occupation difference $n_+-n_-$ of the split
levels by $n_+-n_-\simeq n_1-n_2$. Thereby,
\bee n_i(t) &=&  e^{-\int_0^tds\gamma_i(s)} \Big\{ n_i(0) \nonumber \\
&& \,+ \int_0^tds\gamma_i(s)
e^{\int_0^sds'\gamma_i(s')} \tanh(E_i(s)/(2k_BT)) \Big\} \nonumber
\eee
are the occupation number of isolated TS during an adiabatic field sweep with
constant sweep rate and with the initial occupation difference $n_i(0)$. 
For an isolated TS typical experimental bias field sweeps are adiabatic if
$\Delta_0\ge\sqrt{(\vec{p}\vec{F}_{1}/k_B)(h/t_R)}\simeq 400$~nK with Planck's
constant $h$. The flip-flop dynamics of a RP only occurs when the formation of
the RP is adiabatic as well and, thus, the two TSs, which form a RP, must
fullfil $\Delta_{0p}\ge\sqrt{(\vec{p}\vec{F}_{1}/k_B)(h/t_R)}\simeq 400$~nK. 
The relaxation rate $\gamma_i(s)$ of a TS is given by the one phonon rate
\begin{equation} \gamma_i(s) \,=\, \tau^{-1}(s) \,=\,
  \gamma_0\Delta_0^2E(s)\coth\left( \frac{E(s)}{2k_{\rm B}T} \right) 
\end{equation}
with $\gamma_0=(1/c_l^5+2/c_t^5) B^2/(2\pi\rho\hbar^4)$ where $c_{l,t}$ is the
longitudinal and transversal speed of sound, $\rho$ is the mass density of the
glass and $B$ the strain coupling constant~\cite{Jaeck72}. 
The time dependence is given by the time variation of the asymmetry energy due
to the bias field. We neglected additional
relaxation processes which might emerge from the interaction between
the TSs \cite{BurinRelax94} since their relevance is still a matter of debate. 
Including them will change quantitative results but not
the qualitative picture evolved.   

Statistically a RP $A$ with tunneling splittings
$\Delta_{01A}$ and $\Delta_{02A}$ has a
partner RP $B$ whose tunneling splittings $\Delta_{01B}$ and
$\Delta_{02B}$ are chosen to match $\Delta_{01B}=\Delta_{02A}$ and
$\Delta_{02B}=\Delta_{01A}$. The initial asymmetries have to obey
$\Delta_{1B}=\Delta_{1A}$ and $\Delta_{2B}=\Delta_{2A}$ in order that both RP
are formed at $F=F_{RP}$.
According to Eq. (\ref{eq3}) the responses of RPs $A$ and $B$ differ only
by their occupation numbers and we obtain for the combined response of both RPs
\[ \chi' \,\simeq\, \chi_0
\{(n_{1A}-n_{2B})+(n_{1B}-n_{2A})\} \; ,
\]
where all factors except the occupation numbers are included in $\chi_0$ and
the occupation numbers are considered at $F=F_{RP}$ when the RPs are formed. 
Each of the occupation differences $n_{1A}-n_{2B}$ and $n_{1B}-n_{2A}$ only
depend on a single tunneling element. Thus, we can discuss these two RPs 
as if we had two RPs each formed by two TSs with the same energy and
the same tunneling splitting. 

\begin{figure}[t]
\vspace*{5mm}\epsfig{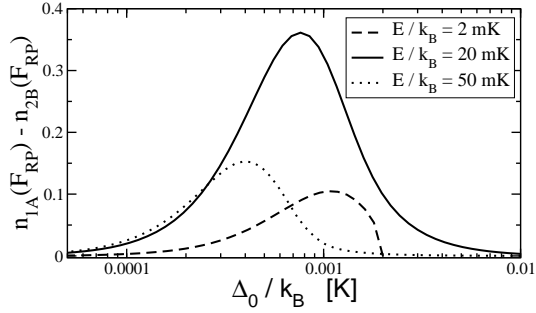}
\caption{\label{fig3} The occupation difference of $n_{1A}-n_{2B}$ is plotted
  versus the tunneling splitting for a temperature $T=10$~mK and
  $\vec{p}\vec{F}_{RP}/k_B=3$~K.} 
\end{figure}

To obtain a substantial dielectric response the RP must be formed with an
occupation difference close to one. If both tunneling elements are the same
the only difference between TS 1 and TS 2 (compare Fig. \ref{fig2}) results
from the bias field sweep; 
the energy splitting of TS 1 is only decreasing until the RP
is formed whereas the energy splitting of TS 2 goes through a minimum. 
Accordingly, we expect a maximum value for the occupation difference of TSs
whose relaxation time is similar to the time the energy splitting of TS 2 is
smaller then $k_{\rm B}T$. 
Note that the relevant RPs are formed by thermal TSs $E\simeq 2k_{\rm B}T$.  
In Fig. \ref{fig3} we plotted
$n_{1A}-n_{2B}$ as a function of $\Delta_0$ for various energy
splittings $E$.

Numerically we find a maximum for
$n_{1A}-n_{2B}$ for a tunneling splitting $\Delta_0^\star$ with 
\[ \tau(\Delta_0^\star(E),E) \,=\, \frac{2k_BT}{\vec{p}\vec{F}_{max}}
\cdot t_{R} 
\]
The full line in Fig. \ref{fig3} represents $n_{1A}-n_{2B}$
for an 
energy splitting $E=2k_BT$ and the dotted and dashed line have $E\gg 2k_BT$
and $E\ll 2k_BT$ respectively. RPs with $E=2k_BT$ contribute maximally and we
define
\be\label{eq99} \Delta_0^\star \,=\,\Delta_0^\star(E=2k_BT) \,=\,
    \sqrt{\frac{\vec{p}\vec{F}_{max}}{t_{R}}\cdot\frac{1}{2\gamma_0 T^2}}
    \; .
\ee
Note that the adiabatic assumption for the RP restricts the tunneling splitting
of the slow relaxing TS to values $\Delta_0>90~\mu$K where we used
$\gamma_0=10^8~$K$^3$s$^{-1}$, $E_{\rm max}=1$~K and
$\overline{J}(T)=100~k_{\rm B}\mu$K at $T=10$~mK. Rosenberg et
al. \cite{Danna02} find a minimal tunneling splitting $\Delta_{0\rm
  min}=250~\mu$K justifying the adiabatic approximation.

To estimate the sudden increase of the dielectric constant
$\delta\varepsilon'^\star/\varepsilon'$ when the field
reaches the maximum we have to count all RPs formed by one TS
with $\Delta_0\simeq\Delta_0^\star$ and a second TS with
$\Delta_0\lesssim\Delta_0^\star$. The values for $n_{1A}-n_{2B}$ and
$n_{1B}-n_{2A}$ are taken from Fig. \ref{fig3} 
and we obtain for the change of the dielectric constant
\be\label{eq11} \frac{\delta\varepsilon'^\star}{\varepsilon'} \,=\,
N_{RP}\overline{\chi}_{RP} \,\simeq\, 1.6\cdot P_0p^2 
\frac{k_BT}{E_{max}}
\ln\left(\frac{\Delta_0^\star}{\Delta_{0min}}\right)
\ee
with the average dielectric response of a contributing RP
\[
\overline{\chi}_{RP} \,=\, \frac{2p^2}{\overline{J}(2k_BT)} 
\frac{1}{2}(n_{1A}-n_{2B})_{max}
\]
and the number of contributing RPs 
\begin{eqnarray}
N_{RP} &=& \int_{k_BT}^{3K_BT}\hspace*{-0.7cm}dE
\int_{\Delta_0^\star/2}^{3\Delta_0^\star/2}\hspace*{-0.75cm}d\Delta_0
\int_{E-\overline{J}}^{E+\overline{J}}\hspace*{-0.65cm}dE'
\int_{\Delta_{0min}}^{\Delta_0^\star}\hspace*{-0.7cm}d\Delta'_0
\frac{P_0}{\Delta_0}\frac{P_0}{(P_0E_{max})\Delta'_0} \nonumber\\
&=& 2\frac{P_0\overline{J}(2k_BT)}{P_0E_{max}} P_02k_BT \ln(3)
\ln\left(\frac{\Delta_0^\star}{\Delta_{0min}}\right) \nonumber 
\end{eqnarray}
%
%
where we approximated $P(E,\Delta_0)\simeq P_0/\Delta_0$ since $\Delta_0\ll E$
for the relevant tunneling splittings. We estimated the number of TSs as
$P_0E_{max}$.
Eq. (\ref{eq11}) predicts a linear increase up to a maximum and then a
linear decrease of $\delta\varepsilon'^\star/\varepsilon'$ with increasing
temperature. Eq. (\ref{eq11}) further predicts a logarithmic dependence
on the sweep rate. Both predictions are in partial agreement with
experiments and we refer to Rosenberg et al. \cite{Danna02} for a detailed
comparison. 

Qualitatively the dynamics of driven RPs explain the experiments except for
the initial sudden decrease. The response due to the driven RPs is always
positive. Thus, our model predicts a sudden increase in $\varepsilon'$ at the
beginning of the first sweep and another such an increase when the field
starts decreasing.  
The experimental observed initial decrease might be explained by
considering the AC measuring field. Typically, the energy shift by the AC
field $\vec{p}\vec{F}_{AC}$ is of the order of $0.24-2.4$~mK~$k_{\rm B}$ 
\cite{Danna02} by assuming dipole moments of $1$ Debye. At temperatures
between $5-20$~mK one might accordingly expect the AC measuring field to drive
the TSs substantially. Thus, the 
DC field provides only a 'second' sweep for the RPs formed near zero field. 
As soon as the DC field exceeds the AC field all RPs are formed the first time
yielding a sudden decrease at the beginning of the DC field sweep as
experimental seen.
The typical energy splitting of RPs is small compared to the energy shift
induced by the experimental AC field. Therefore the AC field influences the
flip-flop dynamics of the RPs. A future
investigation should take into account the AC field beyond linear response but
this is beyond the scope of the present letter. 

In conclusion, we have shown that a first sweep of an electric bias field
prepares resonant pairs of tunneling systems, which are temporarily formed
during the sweep, in metastable states.  
Resonant pairs in these metastable states significantly contribute to the
dielectric constant leading to a higher dielectric constant in subsequent
sweeps, and thus yielding a memory for previously applied fields.

\begin{acknowledgments}
I wish to thank D. Rosenberg, S. Ludwig and D.D. Osheroff for helpful
discussions. I also wish to thank the DOE grant DE-FG03-90ER45435-M012 and the
Alexander-von-Humboldt foundation for support.
\end{acknowledgments}

\end{document}